\documentclass[12pt]{article}

\usepackage{graphicx}
\usepackage{subfigure}
\usepackage{amsmath}
\usepackage{amssymb}

\usepackage{lipsum}
\usepackage{blindtext}

\makeatletter
\newenvironment{tablehere}
{\def\@captype{table}}
{}

\makeatother

\begin{document}
\begin{center}
\begin{large}
\title\\{ \textbf{ An Improved Analysis of Masses and Decay constants of heavy flavour mesons within Variational approach}}\\\
\end{large}

\author\

\textbf{$ Jugal\;Lahkar^{\emph{a}}\footnotemark \:\:,\: D\:K\:Choudhury^{\emph{a,b,c}}and\: B\:J\:Hazarika^{\emph{b}}$ } \\

\footnotetext{Corresponding author. e-mail :  \emph{neetju77@gmail.com}}
\textbf{a}. Dept.of Physics, Gauhati University, Guwahati-781014, India.\\
\textbf{b}. Centre of theoretical Physics, Pandu College, Guwahati-781012, India.\\
\textbf{c}.Physics academy of North-East(PANE),Guwahati-781014,India.\\

\begin{abstract}
We employ the variational method to study the properties such as masses,decay constants,Oscillation frequency and Branching ratios of leptonic decays of heavy flavour mesons with linear cum coulomb Cornell potential. Gaussian function, Coulomb wave function and Airy function are taken as the trial wave-function of variational method in this study.   Our analysis suggests that Gaussian trial wave-function provides results which are in close proximity with the experimental results.  We also make a comparison with the results from QCD Sum rules and lattice QCD ,as well as with recent PDG data .
\end{abstract}
\end{center}
Key words : Quantum Chromo Dynamics, Decay Constant, meson mass. \\\
PACS Nos. : 12.39.-x , 12.39.Jh , 12.39.Pn.\\

\section{Introduction:}\rm
 In recent years,we have been pursuing a QCD based model [1,2,3,4,5]using the linear cum Coulomb Cornell potential  and reported the results of various static and dynamic properties of  heavy  flavour mesons with considerable theoretical success.  The model uses the perturbation method of quantum mechanics [6] where  either of the two pieces of potential (linear or coulomb) is taken as parent and perturbation successively and chosen the better option by comparing with data. However such approach is fraught with inherent limitation:
there are intermediate range of  inter quark separation where both short range  Coulomb and the long range linear parts of the potential  are equally importance and  preference of one above the  other as parent or perturbation  makes  no sense.  It is therefore,instructive to go back to the traditional Variational method[7],[8],[9],[10] and see if a proper trial wave-function can effectively generate the Coulomb and Linear effect without assuming such divide between the long range and the short range effects.

 Another shortcoming of the previous approach with perturbation method is the proper incorporation of relativistic effect in the motion of the light quark in Heavy-Light mesons.  In this context,earlier,a Dirac modification factor $(\frac{r}{a_0})^{-\epsilon}$ [2,4],was introduced as overall relativistic correction,in analogy with QED[11].However,as will be discussed below,such effect does not conform to the the necessity of positivity of mass of heavy-light mesons.  We therefore abandon this modification factor,instead we modify the bare quark mass in meson Hamiltonian by introducing a term $\frac{p^2}{2m}$, consistent with the Hamiltonian used recently by Hwang,Kim and Namgung[7].\\

 Application of variational method in Heavy quark physics was first started by Hwang etal.[7] using the linear cum coulomb potential ,which was later successfully applied by Rai et.al[10] to find masses and decay constants of heavy flavour mesons using a power law potential($\sim \frac{-A}{r}+br^{\nu}$).  While the former[7] calculated only masses and decay constants of pseudo-scalar mesons and the later[10] calculated also the mass difference between pseudo-scalar and vector mesons using the method. More recently,the variational method is applied to heavy flavour physics by Vega and Flores using super-symmetric potential[8].\\

  The above discussion suggests the relevance of variational method in the present day quark dynamics.This has motivated us to study the QCD based potential model  afresh using variational method.  \\

   The manuscript is arranged as below :\\
     In section 2 we provide the formalism where we use variational method  with Gaussian, Coulomb and Airy trial function.In the same section,we also discuss the relativistic correction for the light quark/anti-quark present in heavy-light meson.  In section 3 we provide the results of mass,decay constant,mass difference of pseudo-scalar and vector mesons,ratios of decay constants.Using the masses and decay constants we then calculate the oscillation frequencies  of neutral B mesons and branching ratios of leptonic decays of charged mesons. A comparison is also made with QCD sum rules,Lattice QCD and experimental results.  In section 4 we summarize Conclusion. 

\section{Formalism:}\rm

\subsection{Variational approach with Gaussian trial wave-function:}\rm

We apply a variational method which is similar to[7], considering the trial wave-function to be of Gaussian form as,
\begin{equation}
\psi(r)={(\frac{\alpha}{\sqrt{\pi}})^{\frac{3}{2}}}e^{-\frac{\alpha^2 r^2}{2}}
\end{equation}
where,$\alpha$ is the variational parameter.  We choose the $Q\bar{Q}$ potential as,
\begin{equation}
V(r)=-\frac{4\alpha_s}{3r}+br
\end{equation}
where $\alpha_s$ is the strong coupling constant and $b$ is the standard confinement parameter($b\sim 0.183 GeV^{2}$).  This is the general form of Cornell potential which considers both the properties of quark interaction-asymptotic freedom and confinement, Following variational scheme,the ground state energy is given by,
\begin{equation}
E(\alpha)=\langle\psi\vert H \vert\psi\rangle
\end{equation}

   Now,using the trial wave-function we obtain the expectation values of each term in the Hamiltonian,
\begin{equation}
<-\frac{\nabla^2}{2\mu}> = -\frac{3\mu^2\alpha^2}{4}
\end{equation}
\begin{equation}
<\frac{-4\alpha_s}{3r}> = -\frac{2\mu^3\alpha A}{\sqrt{\pi}}
\end{equation}
\begin{equation}
<br > = \frac{2\mu^3 b}{\sqrt{\pi}\alpha}
\end{equation} 
Therefore,adding all the three equations above we get,
\begin{equation}
E(\alpha)=-\frac{3\mu^2\alpha^2}{4}-\frac{2\mu^3\alpha A}{\sqrt{\pi}}+\frac{2\mu^3 b}{\sqrt{\pi}\alpha}
\end{equation}
where,$A=\frac{4\alpha_s}{3}$ .  Now,by minimising $E(\alpha)$ with respect to $\alpha$ we can find the variational parameter $\alpha$ for different heavy flavoured mesons.  The minimization condition to find the expectation value of Hamiltonian as,
\begin{equation}
\frac{dE(\alpha)}{d\alpha}=0                             
\end{equation}
at $\alpha={\alpha'}$.  

This equation is solved by using $Mathematica7$ and we find the variational parameter for different Heavy Flavour mesons which is shown in $Table.1$.

\begin{tablehere}\scriptsize
\begin{center}
\caption{Variational parameter for different Heavy Flavour mesons:}
\begin{tabular}{|c|c|}
 \hline
          
  Heavy Flavour Mesons& Variational parameter($\alpha')$                  \\\hline
$D{(c\overline{u}/\overline{c}d)}$&0.3087             \\\hline
$D{(c\overline{s})}$&0.310                      \\\hline

$B{(u\overline{b}/d\overline{b})}$ & 0.301 \\\hline
    $B_s{(s\overline{b})}$&0.3473 \\\hline
    $B{(\overline{b}c)}$ &0.537  \\\hline
\end{tabular}
\end{center}
\end{tablehere}

\subsection{ Variational approach with Coulombic trial wave-function:}\rm

Let us now consider the trial wave-function to be,
\begin{equation}
\psi(r)=\frac{(\mu \alpha')^{\frac{3}{2}}}{\sqrt{\pi}}e^{-\mu \alpha' r}
\end{equation}
where,$\alpha'$ is the variational parameter.
Now,with the potential as given in equation(2)we calculate the expectation value of the hamiltonian,
\begin{equation}
E(\alpha')=<\psi\mid H\mid \psi> = \frac{1}{2}\mu \alpha'^2-A\mu \alpha'+\frac{3b}{2\mu \alpha'}
\end{equation}
where,$A=\frac{4\alpha_s}{3}$,and $\alpha_s$ is the strong coupling constant.Now,minimising ,$\frac{dE}{d\alpha'}=0$ ,we get,
\begin{eqnarray}
\alpha'^3-A\alpha'^2-\frac{3b}{2\mu^2}=0
\end{eqnarray}
This equation is solved by using $Mathematica7$ and we find the variational parameter for different Heavy Flavour mesons which is shown in $Table.2$.

\begin{tablehere}\scriptsize
\begin{center}
\caption{Variational parameter for different Heavy Flavour mesons:}
\begin{tabular}{|c|c|}
 \hline
          
  Mesons&$\alpha'$                  \\\hline
$D{(c\overline{u}/\overline{c}d)}$&1.7285               \\\hline
$D{(c\overline{s})}$&1.4642                        \\\hline

$B{(u\overline{b}/d\overline{b})}$ & 1.51164 \\\hline
    $B_s{(s\overline{b})}$&1.230 \\\hline
    $B{(\overline{b}c)}$ &0.6978  \\\hline
\end{tabular}
\end{center}
\end{tablehere}
\subsection{Variational approach with Airy Trial wave-function:}
It is well known that the solution of Schrodinger equation with linear potential gives a wave-function that contains Airy function.  We therefore consider the trial wave-function as Airy function[12],[13],
\begin{equation}
\psi(r)=\frac{N}{2\sqrt{\pi}r}A_{i}[(2 \mu b')r^{\frac{1}{3}}+\varrho _{0n}]
\end{equation}
Here,$b'$ is the variational parameter and $\varrho_{0n}$ are the zeroes of airy function such that $A_{i}[\varrho_{0n}]=0$,and is given as[14]:
\begin{equation}
\varrho_{0n}=-[\frac{3\pi(4n-1)}{8}]^{\frac{2}{3}}
\end{equation}
For different S states few zeroes of the Airy function is listed below:\\\\

\begin{tablehere}\scriptsize
\begin{center}
\caption{Zeroes of Airy function for different S-states:}
\begin{tabular}{|c|c|}
 \hline
          
  States&$\varrho_{0n}$                  \\\hline
1s(n=1,l=0)&-2.3194                     \\\hline
2s(n=2,l=0)& -4.083                         \\\hline
3s(n=3,l=0)& -5.5182                      \\\hline
4s(n=4,l=0)&-6.782                           \\\hline
\end{tabular}
\end{center}
\end{tablehere}

It is worthwhile to mention that  Airy function  is an infinite series in itself,given as:
\begin{equation}
A_i[\varrho]=a_0[1+\frac{\varrho ^3}{3!}+\frac{\varrho ^6}{6!}+\frac{\varrho ^9}{9!}+.....]-b_0[\varrho+\frac{\varrho ^4}{4!}+\frac{\varrho ^7}{7!}+\frac{\varrho ^{10}}{10!}+.....]
\end{equation}
with,$a_0=0.3550281$ and $b_0=0.2588194$.\\
  Now,the normalization condition is,
\begin{equation}
\int_{0}^{\infty}4\pi r^2 \psi^*\psi dr=1
\end{equation}
Substituting the wave-function from equation (12) we get,
\begin{equation}
N=\frac{1}{[\int_{0}^{\infty}A_i^2[(2\mu b')^{\frac{1}{3}}r-2.3194]^{\frac{1}{2}}}
\end{equation}
This can be easily calculated in Mathematica.7 .  The corresponding energies are[13],[14],
\begin{equation}
E_n=-[\frac{b'^2}{2\mu}]^{\frac{1}{3}}\varrho_{0n}
\end{equation}
While dealing with the Airy function as the trial wave-function of variational method with the Cornell potential,the main problem is that the wave-function has got a singularity at $r=0$. The presence of singularity in a wave-function is not new and in QED also such singularities appear[11],[15],[16].  
Therefore,to calculate the wave-function at the origin,we follow a method valid for S-wave as[17].  In this method,the wave-function at the origin is found from the condition,$\vert\psi(0)\vert^2=\frac{\mu}{2\pi}\langle\frac{d V}{d r}\rangle$[17].  We find  the variational parameter $b'$ as,

\begin{tablehere}\scriptsize
\begin{center}
\caption{Variational parameter for Airy trial function:}
\begin{tabular}{|c|c|}
 \hline
          
  Mesons&$b'$                 \\\hline
$D{(c\overline{u}/\overline{c}d)}$&2.050             \\\hline
$D{(c\overline{s})}$&1.597         \\\hline

$B{(u\overline{b}/d\overline{b})}$ & 1.7269 \\\hline
    $B_s{(s\overline{b})}$&1.2709 \\\hline
    $B{(\overline{b}c)}$ &0.558  \\\hline
\end{tabular}
\end{center}
\end{tablehere}
\subsection{Relativistic effect in light quark:}
To study Heavy-light meson in any potential model one needs to incorporate the relativistic effect in the light quark. In some of our previous works [1,2,3,4,5],the relativistic correction to the wave-function was made by introducing a Dirac modification term $(\frac{r}{a_0})^{-\epsilon}$,where $\epsilon=1-\sqrt{1-(\frac{4\alpha_s}{3})^2}$ in analogy with QED [11],[15]. As noted in Sakurai[11] the term $(\frac{r}{a_0})^{-\epsilon}$ is essentially unity except for a very short distance.  When $r\rightarrow0$,the wave function develops a singularity. In the previous works[1-5],a prescription is provided through the the regularization of the wave-function at the origin with suitable a cut-off $r_0$,which was based on QED analogy.  The modification is,
\begin{eqnarray}
r'=r+r_0\\
r_0\sim a_0e^{-\frac{1}{\epsilon}}
\end{eqnarray}
where,$\epsilon=1-\sqrt{1-(\frac{4\alpha_s}{3})^2}$ is the Dirac modification factor and $a_0=\frac{3}{4\mu\alpha_s}$.  But,the problem with such regularization is that ,for pseudo-scalar mesons,the masses losses positivity. The mass formula for pseudo-scalar mesons,\\
\begin{equation}
M_P = m_Q +m_{\overline{Q}}-\frac{8\pi\alpha_s}{3m_Qm_{\bar{Q}}}{\mid\psi(0)\mid}^2
\end{equation}

 Therefore from the positivity of mass we obtain the corresponding lower bound on the cut-off parameter $r_0$ as follows,
\begin{eqnarray}
m_Q+m_{\bar{Q}}\geq\frac{8\pi\alpha_s}{3m_Qm_{\bar{Q}}}\vert\psi(0)\vert ^2
\end{eqnarray}
From this inequality,with suitable cut-off $r_0$ in the airy trial wave-function,we calculate the lower bound on $r_0$ from positivity of mass,which are shown below for various Heavy Flavour mesons.  As an illustration,experimental value of mass of the $D$ meson $(1.869GeV)$ will yield a value of $r_0 \sim 19GeV^{-1}$ ,which exceeds the size of meson itself.  From the table.4 it is clear that the regularization of wave-function with QED analogy as depicted in[2,4] fails from the prospective of positivity of mass. The positivity of mass of pseudo-scalar mesons yields a regularization length which exceeds that given by QED analogy of H-atom.  This is a distinctive feature of Heavy flavour mesons of QCD compared to H-atom of QED. 

\begin{tablehere}\scriptsize
\begin{center}
\caption{Lower bound on cut-off $r_0$ from positivity of mass(in $GeV^{-1}$):}
\begin{tabular}{|c|c|c|}
 \hline
          
  Heavy Flavour Mesons& $r_0$(from positivity of mass)&   $r_0[2,4]$               \\\hline
$D{(c\overline{u}/\overline{c}d)}$&0.089& 0.0073           \\\hline
$D{(c\overline{s})}$&0.037&0.0055                      \\\hline

$B{(u\overline{b}/d\overline{b})}$ & 0.0035 &$1.452\times10^{-9}$\\\hline
    $B_s{(s\overline{b})}$&0.0033& $1.038\times10^{-9}$\\\hline
    $B{(\overline{b}c)}$ &0.002 &$3.872\times10^{-9}$ \\\hline
\end{tabular}
\end{center}
\end{tablehere}

 Therefore,the relativistic effect in the light quark is introduced as in reference[6]. In the works [7],[10],the light quark is considered relativistically with the Hamiltonian as $H=M+\frac{p^2}{2m}+\sqrt{p^2+m^2}+V(r)$,where,M is the mass of Heavy quark and m is the mass of light quark. In the present work,we just use a simplified variant of [7] and take only first order corrections of the relativistic effect as $E=m+\frac{p^2}{2m}$ in the light quark.\\

\subsection{Masses of Heavy Flavour mesons:}
Pseudo-scalar meson mass can be computed from the following relation [2],[16],[18]:
\begin{eqnarray}
M_P = M+m+\frac{p^2}{2m}+ \triangle E  
\end{eqnarray}

Here,$M$ and $m$ are the masses of Heavy quark/anti-quark and light quark/anti-quark respectively and we have considered 1st order relativistic correction to the light quark/anti-quark. The energy shift of mass splitting due to spin interaction in the perturbation
theory reads[1,2],[6],[16],

\begin{equation}
\triangle E=\frac{32\pi\alpha_s}{9Mm}{S_Q.S_{\bar{Q}}}{\mid\psi(0)\mid}^2
\end{equation}
For pseudo-scalar mesons,$S_Q.S_{\bar{Q}}=-\frac{3}{4}$,so pseudo-scalar meson masses can be expressed as,
\begin{equation}
M_P = M+m+\langle -\frac{\nabla^2}{2m}\rangle-\frac{8\pi\alpha_s}{3Mm}{\mid\psi(0)\mid}^2
\end{equation}
Similarly,for vector mesons $S_Q.S_{\bar{Q}}=\frac{1}{4}$, so,
\begin{equation}
M_V =  M+m+\langle -\frac{\nabla^2}{2m}\rangle+\frac{8\pi\alpha_s}{9Mm}{\mid\psi(0)\mid}^2
\end{equation}
This particular aspect was overlooked in references [2-5],[13].\\\\
 
\subsection{Decay constants of heavy flavoured mesons:}\rm
For pseudo-scalar mesons,the decay constant $f_p$ is related to the ground state wave function at the origin $\psi(0)$ according to the Van-Royen-Weisskopf formula[19], in the non-relativistic limit as,
\begin{equation}
f_p=\sqrt{\frac{12{\mid\psi(0)\mid}^2}{M_p}} 
\end{equation}
where,$M_p$ is the mass of pseudo-scalar meson .  Now with QCD correction factor [2] it can be written as,

\begin{equation}
f_p=\sqrt{\frac{12{\mid\psi(0)\mid}^2}{M_p}{\bar{C}^2}}
\end{equation}

With,
\begin{equation}
\bar{C}^2=1-\frac{\alpha_s}{\pi}[2-\frac{m_Q-m_{\bar{Q}}}{m_Q+m_{\bar{Q}}}ln\frac{m_Q}{m_{\bar{Q}}}]
\end{equation}

   Again,the ratios of pseudo-scalar decay constants eg.for $B_S$ and $D_s$ meson can be expressed as,
\begin{equation}
\dfrac{f_{B_s}}{f_{D_S}}=\sqrt{\dfrac{M_{D_S}}{M_{B_S}}}\dfrac{\psi_{B_S}(0)}{\psi_{D_S}(0)}
\end{equation}

\subsection{Mass difference of vector and Pseudo-scalar mesons:}\rm
The mass difference between the Pseudo-scalar and vector meson is given by[10],[20],
\begin{equation}
M_{{(Q\bar{Q})}^*} -M_{(Q\bar{Q})}=\frac{8\pi A}{3m_Qm_{\bar{Q}}}{\mid\psi_{Q\bar{Q}}(0)\mid}^2
\end{equation}
where $m_Q$ is the mass of heavy quark and $m_{\bar{Q}}$ is the mass of antiquark.
This is attributed to the hyperfine interaction and $A=\frac{4\alpha_s}{3}$ where $\alpha_s$ is the strong coupling constant.\\

\subsection{Oscillation frequency:}
It has been well established that the $B_D$ and $B_S$ meson mix with their antiparticles by means of Box diagram and involves exchange of $W$ bosons and $u,c,t$ quarks which leads to oscillation between mass eigenstates[21],[22],[23.  The oscillation is parametrized by mixing mass parameter $\bigtriangleup m$ given by,
\begin{equation}
\Delta m_B =\frac{G_F^2m_t^2M_{B_q}f_{B_q}^2}{8\pi}g{(x_t)}\eta _t\mid V_{tq}^*V_{tb}\mid ^2 B
\end{equation}
Where,$\eta _t$ is the gluonic correction to oscillation(=0.55[24]) and $B$ is the bag parameter(=1.34[24]) and the parameter $g(x_t)$ is given as [25],
\begin{equation}
g(x_t)=\frac{1}{4}+\frac{9}{4(1-x_t)}-\frac{3}{2(1-x_t)^2}-\frac{3x_t^2}{2(1-x_t)^3}
\end{equation}
and,$x_t=\frac{m_t^2}{M_W^2}$. From data of Particle data group[26],
$m_t=174GeV$,$M_W=80.403GeV$,$\mid V_{tb}\mid=1$,$\mid V_{td}\mid=0.0074$,$\mid V_{ts}\mid=0.04$.

\subsection{Leptonic decay width and Branching ratio:}
It is also well known that charged mesons $(\pi^{\pm},K^{\pm},D^{\pm},D_{S}^{\pm},B^{\pm})$ can decay to a charged lepton pair,when they annihilate via a virtual $W^{\pm}$ boson.  Purely leptonic decays are rare, but there are clear experimental signatures because of the presence of highly energetic lepton in the final state.  Absence of hadrons in the final state indicates that the theoretical predictions are very clean. The partial decay width for the process is given by[22],
\begin{equation}
\Gamma(P\rightarrow l\nu)=\frac{G_F^2}{8\pi}f_P^2M_Pm_l^2(1-\frac{m_l^2}{M_p^2})^2\vert V_{fg}\vert ^2
\end{equation}
With the computed masses and decay constants,the leptonic decay widths for separate lepton channel $m_{l=\mu,\tau,e}$can be easily calculated.  Here,$ V_{fg}$ is the CKM matrix element for quark flavours $f$ and $g$ ,also,  $G_F , P , f_P , M_P ,m_l$ denote the Fermi constant, generic pseudo-scalar(PS)meson, PS-meson weak-decay constant, PS-meson mass and lepton mass respectively.  The branching ratio of heavy flavour mesons is calculated by using relation,
\begin{equation}
\textit{B}=\tau_P\Gamma(P\rightarrow l\nu)
\end{equation}
Here,$\tau_P$ is the lifetime of pseudo-scalar mesons.  For calculation we take the world average values reported by particle data group[26] as , $\tau_{D} =1.04ps$,$\tau_{D_{s}} =0.5 ps$,$\tau_B =1.63 ps$ and $\mid V_{cd} \mid=0.230$,$\mid V_{cs} \mid=1.023$,$\mid V_{ub} \mid=3.89\times10^{-3}$ . 

\section{Results:}\rm
\subsection{Masses:}
With the formalism developed in section $2$ ,we calculate the masses  of some  Heavy Flavour mesons,which are shown in $Table.6$ . The input parameters are $m_{u/d}=0.336Gev$ ,$m_b=4.95GeV$, $m_c=1.55GeV$, $m_s=0.483GeV$ and $b=0.183GeV^2$[4],[26],also we take $\alpha_s=0.39$ for C-scale and $\alpha_s=0.22$ for b-scale [4].  We  make a comprehensive comparison of our results with lattice results[27],QCD sum rule[28],other models[4],[9] and present experimental results. Our result agrees well with the present results of  lattice QCD,QCD sum rules and experimental data.  The difference of the predictions of our model with the more advanced approaches such as lattice QCD[27] and QCD sum rules[28] are insignificant.  As an illustration,from Table.6,the predicted  mass of D and B meson with Gaussian trial wave-function are ($1.94$GeV and $5.35 $GeV),which are quite close to the lattice results ($1.885$GeV and $5.283$ GeV)and QCD sum rule results ($1.87$ GeVand$5.283$GeV).  The pattern is similar to other mesons.  Similarly,the mass difference between the pseudo-scalar and the vector mesons are given in $Table.9$ and compared with lattice results.  Here too,our results agrees with lattice results.

\subsection{Decay constants:}
The decay constants of Heavy flavoured mesons are calculated using equation(27) and are shown in Table.7,where also comparison is made with the results of lattice QCD,QCD sum rules and experimental data.  Here too,the agreement of the predictions of our model with a Gaussian trial wave-function with that of lattice QCD[29],[30],[31] and QCD sum rules[32] are good.  As an illustration,the decay constant of B meson is ($0.198$GeV),which is close to lattice result($0.218$GeV)and QCD sum rule result($0.193$GeV).  Again in non relativistic case,$\frac{f_B}{f_D}\simeq \sqrt{\frac{M_D}{M_B}}=0.59$,but using relativistic correction to the light quark/anti-quark we get, $\frac{f_B}{f_D}=0.70=0.59\times1.18$,i.e.$\frac{f_B}{f_D}$ is enhanced by a factor $1.18$.  This is to be compared with Hwang's work,where the factor is, $1.13,1.31$[6]($\frac{f_B}{f_D}=0.67 or0.77$). In $table.8$ we show the ratios of decay constants and compared with the lattice results.   Also the ratio $\xi=\frac{\sqrt{B_S}f_{B_S}}{\sqrt{B_D}f_{B_D}}\simeq \frac{f_{B_S}}{f_{B_D}}=1.045$ which is well agreement with[22].
\subsection{Oscillation frequencies :}
    The mixing mass parameter $\Delta m$ ,which is connected to the oscillation of neutral mesons is estimated and compared with the results of QCD sum rules,lattice QCD and experimental data,which is shown in table.10.  Here too,our predictions with a Gaussian trial wave-function is in good agreement with lattice results[33],QCD sum rule result[34] and the experimental data[35],[36].
    
\subsection{Branching ratios:}    
     The Branching ratio of different decay channels in the leptonic decays of Heavy Flavour mesons are calculated and shown in $Table.11(Gaussian)$ , $Table.12(Coulomb)$ and $Table.13(Airy)$. Comparison is also done with other model[37] and the experimental results.\\\\
       From all the tables,it can be easily inferred that while exploring static properties of Heavy Flavour mesons variational approach with Gaussian trial wave-function is phenomenologically preferable.

\begin{tablehere}\scriptsize
\begin{center}
\caption{Masses of Heavy Flavoured mesons(in GeV)}
\begin{tabular}{|c|c|c|c|c|c|c|c|c|}
  \hline
          
    Mesons  &  $M_{P}$(Gaussian) & $M_{P}$(Coulomb) & $M_{P}$(Airy)&[16]&  [4] & lattice[27]&Q.sum rule[28]& Exp.Mass   \\
    \hline
    $D{(c\overline{u}/cd)}$   &1.94 &1.606&1.28&1.972& 2.378 &1.885&1.87  & $1.869\pm0.0016$  \\ \hline
    $D{(c\overline{s})}$  & 2.032 &1.739&1.69&2.154 &2.076&1.969&1.97& $1.968\pm0.0033$  \\\hline
    $B{(u\overline{b}/d\overline{b})}$   &5.35&5.11&5.14&5.314 & 5.798&5.283 &5.28&$5.279\pm0.0017$   \\\hline
    $B_s{(s\overline{b})}$&  5.48 &5.40&5.372&5.6  &5.331&5.366&5.37&$5.366\pm0.0024$     \\\hline
    $B{(\overline{b}c)}$ &6.4 &6.38&6.5& &  6.8&6.278  && $6.277\pm0.006$  \\
    \hline
\end{tabular}
\end{center}
\end{tablehere}

\begin{tablehere}\scriptsize
\begin{center}
\caption{Decay constants of Heavy Flavoured mesons(GeV)}
\begin{tabular}{|c|c|c|c|c|c|c|}
  \hline
          
    Mesons & $f_p$(Gaussian)  &  $f_p$(Coulomb)  & $f_P$(Airy)&QCD Sum rules[32] &Lattice& Exp.value  \\
    \hline
    $D{(c\overline{u}/cd)}$  & 0.282 & 0.377& 0.801&$0.206\pm0.002$ &$0.220\pm0.003[18]$&$0.205\pm0.85\pm0.025$ [35,38]     \\\hline
    $D{(c\overline{s})}$   &  0.336 & 0.431 & 0.683& $0.245\pm0.015$&$0.258\pm0.001[31]$&$0.254\pm0.059$[35,38]   \\\hline
    $B{(u\overline{b}/d\overline{b}))}$ & 0.198 &0.264 & 0.764&$0.193\pm0.012$ &$0.218\pm0.005[18]$&$0.198\pm0.014$[39] \\\hline
    $B_s{(s\overline{b})}$&0.207  & 0.238&0.627&$0.232\pm0.018$&$0.228\pm0.010[30]$&$0.237\pm0.017$[39]   \\\hline
    $B{(\overline{b}c)}$ &0.563  &0.59 &0.333& &&0.562[24]    \\
    \hline
\end{tabular}
\end{center}
\end{tablehere}

\begin{tablehere}\scriptsize
\begin{center}
\caption{ Ratios of pseudoscalar decay constants:}
\begin{tabular}{|c|c|c|c|c|c|c|}
  \hline
          
   Ratios & $\frac{f_{B_s}}{f_B}$(Gaussian)& $\frac{f_{B_s}}{f_B}$(Coulomb) & $\frac{f_{B_s}}{f_B}$(Airy) & $\frac{f_{D_S}}{f_D}$ (Gaussian)&$\frac{f_{D_S}}{f_D}$(Coulomb) &$\frac{f_{D_S}}{f_D}$(Airy) \\ \hline
Our model&$ 1.045$&$0.90$ &0.821&$1.19$ &$1.14$& 0.857   \\\hline
Lattice result&$1.16\pm0.06[40]$& &&  $1.188[37]$& &            \\ \hline
QCD Sum rule&$1.20\pm0.03$[41]&&&$1.17\pm0.03$[41]&&           \\\hline
    
\end{tabular}
\end{center}
\end{tablehere}

\begin{tablehere}\scriptsize
\begin{center}
\caption{Mass difference of pseudo-scalar and vector mesons(GeV):}
\begin{tabular}{|c|c|c|c|c|}
   \hline
          
    Mesons & $M_v-M_p(Gaussian)$   &  $M_v-M_p(coulomb)$&$M_v-M_p(Airy)$&$M_v-M_p(Lattice)[38]$  \\
    \hline
    $D{(c\overline{u}/cd)}$  &  0.44   &  0.81&0.66 & 0.067   \\\hline
    $D{(c\overline{s})}$   &  0.312 &0.28 & 0.46& 0.066 \\\hline
    $B{(u\overline{b}/d\overline{b})}$ & 0.038 &0.088& 0.11 &0.034 \\\hline
    $B_s{(s\overline{b})}$&  0.026 & 0.16 &0.081&0.027 \\\hline
    $B{(\overline{b}c)}$ & 0.0305  &0.0305 &0.025&    \\
    \hline
\end{tabular}
\end{center}
\end{tablehere}

\begin{tablehere}
\begin{center}
\caption{ Mixing mass parameter($ps^{-1}$):}
\begin{tabular}{|c|c|c|c|c|c|c|}
  \hline
          
   Meson&$\Delta m_B$(Gaussian)&$\Delta m_B$ (Coulomb)& $\Delta m_B$(Airy)&sum rule&lattice&Exp.value  \\\hline
   $B_D$ &$0.45 $&0.78 &0.27&0.48[34]&0.63[33]&0.5[36]                                           \\\hline
   $B_S$ &$16$ &60&9.3&$>14.6[34]$&19.6[33]&17.76[35]                                    \\\hline
\end{tabular}
\end{center}
\end{tablehere}

\begin{tablehere}\scriptsize
\begin{center}
\caption{ Branching ratio of Heavy Flavour mesons(with Gaussian):}
\begin{tabular}{|c|c|c|c|}
  \hline
 \textbf{Mesons} & $\textit{B}R_{\tau}\times10^{-3}$& $\textit{B}R_{\mu}\times10^{-4}$&  $\textit{B}R_{e^-}\times10^{-8}$            \\\hline
  D     & $2.08$(this work) & $7.6$(this work) & $1.82$(this work)           \\\hline
  [37]& $0.9$   &$6.6$&$1.5$              \\\hline
  Expt.[24]&$<2.1$ &$4.4\pm0.7$&$<8.8$             \\\hline
  \textbf{Mesons} & $\textit{B}R_{\tau}\times10^{-2}$& $\textit{B}R_{\mu}\times10^{-3}$&  $\textit{B}R_{e^-}\times10^{-7}$            \\\hline
  $D_{s}$&$6.94$&$10$&$0.0025$                    \\\hline
  [37]&$8.4$&$7.7$&$1.8$                  \\\hline
  Expt.[24]&$6.6\pm0.6$&$6.2\pm0.6$&$<1.2$              \\\hline
   \textbf{Mesons} & $\textit{B}R_{\tau}\times10^{-4}$& $\textit{B}R_{\mu}\times10^{-6}$&  $\textit{B}R_{e^-}\times10^{-6}$            \\\hline
   B&$1.14$&$0.489$&$0.00014$                     \\\hline
   Expt.[26]&$1.8$&$<1$&$<1.9$                        \\\hline
\end{tabular}
\end{center}
\end{tablehere}
\pagebreak
\begin{tablehere}\scriptsize
\begin{center}
\caption{ Branching ratio of Heavy Flavour mesons(with Coulomb):}
\begin{tabular}{|c|c|c|c|}
  \hline
 \textbf{Mesons} & $\textit{B}R_{\tau}\times10^{-3}$& $\textit{B}R_{\mu}\times10^{-4}$&  $\textit{B}R_{e^-}\times10^{-8}$            \\\hline
  D     & 2.5 & $7.6$ & $1.82$           \\\hline
  [37]& $0.9$   &$6.6$&$1.5$              \\\hline
  Expt.[24]&$<2.1$ &$4.4\pm0.7$&$<8.8$             \\\hline
  \textbf{Mesons} & $\textit{B}R_{\tau}\times10^{-2}$& $\textit{B}R_{\mu}\times10^{-3}$&  $\textit{B}R_{e^-}\times10^{-7}$            \\\hline
  $D_{s}$&$6.94$&$10$&$0.0025$                    \\\hline
  [37]&$8.4$&$7.7$&$1.8$                  \\\hline
  Expt.[24]&$6.6\pm0.6$&$6.2\pm0.6$&$<1.2$              \\\hline
   \textbf{Mesons} & $\textit{B}R_{\tau}\times10^{-4}$& $\textit{B}R_{\mu}\times10^{-6}$&  $\textit{B}R_{e^-}\times10^{-6}$            \\\hline
   B&$1.14$&$0.489$&$0.00014$                     \\\hline
   Expt.[26]&$1.8$&$<1$&$<1.9$                        \\\hline
\end{tabular}
\end{center}
\end{tablehere}
\begin{tablehere}\scriptsize
\begin{center}
\caption{ Branching ratio of Heavy Flavour mesons(with Airy):}
\begin{tabular}{|c|c|c|c|}
  \hline
 \textbf{Mesons} & $\textit{B}R_{\tau}\times10^{-3}$& $\textit{B}R_{\mu}\times10^{-4}$&  $\textit{B}R_{e^-}\times10^{-8}$            \\\hline
  D     & 7.7 & $0.27$ & $61$           \\\hline
  [37]& $0.9$   &$6.6$&$1.5$              \\\hline
  Expt.[24]&$<2.1$ &$4.4\pm0.7$&$<8.8$             \\\hline
  \textbf{Mesons} & $\textit{B}R_{\tau}\times10^{-2}$& $\textit{B}R_{\mu}\times10^{-3}$&  $\textit{B}R_{e^-}\times10^{-7}$            \\\hline
  $D_{s}$&$70$&$2.4$&$0.55$                    \\\hline
  [37]&$8.4$&$7.7$&$1.8$                  \\\hline
  Expt.[24]&$6.6\pm0.6$&$6.2\pm0.6$&$<1.2$              \\\hline
   \textbf{Mesons} & $\textit{B}R_{\tau}\times10^{-4}$& $\textit{B}R_{\mu}\times10^{-6}$&  $\textit{B}R_{e^-}\times10^{-6}$            \\\hline
   B&$3.26$&$0.11$&$0.0026$                     \\\hline
   Expt.[26]&$1.8$&$<1$&$<1.9$                        \\\hline
\end{tabular}
\end{center}
\end{tablehere}

\section{Conclusion:}\rm
We have investigated the static properties of heavy flavour mesons using variational method with Cornell potential in co-ordinate space with three different trial wave-functions. Specifically,we consider the trial wave-functions viz.Gaussian, Coulomb and Airy function. Also,since the light quark of a Heavy-Light meson is relativistic,we have incorporated first order relativistic correction in a minimal way.  The Gaussian wave-function appears to be better choice compared with Coulomb and Airy trial wave-functions.  We have also compared our results with those of lattice QCD and the  QCD sum rules.  Our results with a Gaussian trial wave-function conform with the results of both of them.  In a sense,it is an improvement over our previous perturbative approaches[1-5,13,22], where an arbitrary choice between the linear and the Coulomb part of Cornell potential as parent/perturbation is necessary. \\
  To conclude,the variational method with a Gaussian trial wave-function provides a simple method to study the static and dynamic properties of pseudo-scalar mesons which are close to the corresponding results of the lattice QCD and QCD sum rules.  Such effective H.O. wave function can presumably be  generated by a  $Q\bar{Q}$ potential, which is polynomial in r,  $V(r)=\Sigma_{n=-l}^{n=+l}a_nr^n, with, a_2 \gg a_{l,l\neq2}$,  a feature noticed by Godfrey and Isgur[42] as early as in 1980's.  The present analysis,therefore seems to indicate the relevance of such a simple model based on the Schrodinger equation as far as phenomenology is concerned,in spite of advanced tools like lattice QCD and QCD sum rules available in the current literature.

\paragraph{Acknowledgement:}
\begin{flushleft}
\emph{One of the authors (Jugal Lahkar) acknowledges the financial support of CSIR(New-Delhi,India) in terms of fellowship under Net-Jrf scheme to pursue research work at Gauhati
University, Department of Physics.Also we thank Dr.N.S.Bordoloi of Cotton university,Guwahati,Assam,India,and Dr.S.Roy of Karimganj College,Karimganj,Assam,India for useful discussions.Another one of us(D.K.Choudhury) thanks Prof.M.P.Borah,HOD,Dept.of Physics,G.U.,Guwahati,Assam,India and Dr.Bandana Das,HOD,Pandu College,Guwahati,Assam,India for providing facilities for research work.}
\end{flushleft}


\begin{thebibliography}{99}

\bibitem{}D.K.Choudhury,P.Das,D.D.Goswami,J.K.Sarma,Pramana J of Phys.,vo.44(1995)
\bibitem{kkp}K. K. Pathak and D. K. Choudhury, Chin. Phys. Lett. \textbf{28}, 101201 (2011)

\bibitem{BJHazarika}B.J.Hazarika and D.K.Choudhury,Pramana.J.Phys.,vol75.,no.3,sept.2010
\bibitem{T das}T Das and D.K.Choudhury,Int.J of Mod.Phy A,2016,DOI: 	10.1142/S0217751X1650189X
 \bibitem{}S.Roy,N.S.Bordoloi and D.K.Choudhury,Can. J. Phys. 91(2013)34 
 \bibitem{}Halzen and Martin "Quarks and Leptons",John Wiley and Sons,ISBN:0-471-88741-2,P65
\bibitem{Hwang}D.S.Hwang etal., Phys. Rev. \textbf{D 53}, 4951,1996
\bibitem{Vega}A.Vega and J.Flores,Pramana – J. Phys. (2016) 87:73
DOI 10.1007/s12043-016-1278-7

\bibitem{Rai}N.Devlani and A.K.Rai,Phys.Rev.\textbf{D84},074030,2011
\bibitem{Rai}A.K.Rai,R.H.Parmar and P.C.Vinodkumar,J. Phys. G: Nucl. Part. Phys. 28 (2002) 2275–2282
\bibitem{Sakurai}J.J.Sakurai,Advaced Quantum Mechanics,p129,(Massachusetts,Addison Wesly publishing company.1967)
\bibitem{M.abramowitz}M.Abramoowitz and I.Stegun,"Handbook of Mathematical functions",National Bureau of Standards,Us,1964
\bibitem{BJHazarika}B.J.Hazarika and D.K.Choudhury,Pramana.J.Phys.,Vol78.,No.4,
April2012
\bibitem{Aitchinson}I.J.R.Aitchinson and J.J.Dudek.,Eur.J.phys.,23,605(2002)
\bibitem{Zuber}C Itzkyson and J zuber,Quantum Field theory,p79(International students edition,McGraw Hill,Singapur,1986)
\bibitem{Griffiths}D Griffiths, Introduction to Elementary Particles;John Wiley and Sons,
New york(1987),p158.
\bibitem{Quigg}Quigg C and Rosner J L 1979 Phys. Rep. 56 167

\bibitem{Eichten}E.Eichten etal.,Phys.Rev.D21(1980)203

\bibitem{van}Van Royen R et al., Nuovo Cimento \textbf{50}, (1967)
\bibitem{}K.Igi and S.Ono,Phys.Rev.D,vol32(1985)
\bibitem{Buras}Buras A,Phys.Lett.B566,115(2003)DOI:10.1016/S0370-2693(03)00561-6
\bibitem{kkp}K.K.Pathak,D.K.Choudhury and N.S.Bordoloi,Int. J. Mod. Phys.A Vol. 28 (2013) 1350010 
\bibitem{}D. Ebert, R. N. Faustov, V. O. Galkin, Phys. Rev. D 67, 014027 (2003).
\bibitem{patel}Bhavin Patel and P C Vinodkumar,Chinese Physics C, Vol. 34, No. 9,(2010); arXiv:hep-ph/0908.2212v1(2009).
\bibitem{}Inami,T and Lim,C.S.Prog.Theo.phys.65(1981);ibid.65(1981)1772(E)297

\bibitem{}C. Patrignani and Particle Data Group, 2016 Chinese Phys. C 40 100001 

\bibitem{}R.j.Dowdall etal.,HPQCD Collab.,arxiv:1207.5149v1

\bibitem{}Z.G.Wang,Eur. Phys. J. C75 (2015) 427,DOI:10.1140/epjc/s10052-015-3653-9


\bibitem{} K. C. Bowler et al., (UKQCD Collaboration), hep-lat/0007020.
\bibitem{}Heechang Naetal.,Phys. Rev. D 86 (2012) 034506,DOI:10.1103/PhysRevD.86.034506
\bibitem{}W.Chen etal.,TWQCD Collab.,Phy.Lett.B,736(2014),https://doi.org/10.1016/j.physletb.2014.07.025
\bibitem{}W.Lucha etal.,J. Phys. G: Nucl. Part. Phys. 38 (2011) 105002 (17pp)
\bibitem{}A. Bazavov etal.,Phys.Rev.D93, 113016 (2016)
\bibitem{}C.Gay,Ann.Rev.Nucl.Part.Sci.50:577-641,2000,DOI:10.1146/annurev.nucl.50.1.577
\bibitem{}The LHCb collab.,Arxiv:1304.4741v1[hep-ex]
\bibitem{}The LHCb collab.,Eur.J.of Phys.C(2016)76:412
\bibitem{yang}M. Z. Yang, Eur. Journ. of Phys. \textbf{C 72}, 1880 (2012)
\bibitem{asner}D. Asner et al. (Heavy Flavor Averaging Group), arXiv:\textbf{1010.1589}.
\bibitem{Eisenstein} B. I. Eisenstein et al. (CLEO Collaboration), Phys. Rev.\textbf{ D 78}, 052003 (2008)
\bibitem{}L. Lellouch et al. (UKQCD Coll.), hep-ph/9912322
\bibitem{}T. Huang and C.W. Luo, Phys. Rev. D 53, 5042 (1996)
\bibitem{}Godfrey and Isgur,Phys.Rev.D32,189(1985)



















\end{thebibliography}
\end{document}